\documentclass[letterpaper,12pt]{article}
\usepackage{tabularx} 
\usepackage{amsmath}
\usepackage{amsmath,bm}

\usepackage{graphicx} 
\usepackage[margin=1in,letterpaper]{geometry} 
\usepackage{cite} 
\usepackage[final]{hyperref} 
\hypersetup{
	colorlinks=true,       
	linkcolor=blue,        
	citecolor=blue,        
	filecolor=magenta,     
	urlcolor=blue         
}
\usepackage{amsfonts}

\usepackage{bbold}

\begin{document}

\title{Optical activity in weakly coupled nonorods}

\author{M. A. Kuntman and E. Kuntman\\ \footnotesize{Baran Sitesi F-6, 52200 Ordu, Turkey}}

\maketitle

\begin{abstract}
 We  introduce a matrix method and we derive a formula for phase retardation effects in plasmonic systems. We analyze the circular dichroic response (CD) of two orthogonal Au nanorods in detail and  we show that, although, theoretically, circular dichroism for forward scattering is directly proportional to the dipole-dipole interaction between the particles, CD response of the system can be much greater in weak coupling due to the trade off between two different types of phases.

\end{abstract} 

\section{Introduction}

Achiral and chiral configurations of coupled plasmonic nanorods manifest circular polarization effects due to the phase difference between the light scattered from different parts of the system. In the Born-Kuhn coupled oscillator model and in its plasmonic versions \cite{Khun,Schaferling,Yin,Pakizeh} optical activity emerges as a result of the coupling (dipole-dipole interaction) between the particles.

For simple nanosystems, it is usually sufficient to investigate the system’s behavior under a single
polarization excitation. However, for coupled nanoparticles in three-dimensional space,
it is often necessary to study different excitation polarizations and
matrix methods comprising all possible modes become important \cite{PRB,Thesis,MDPI}. In this note, we examine
the optical response of coupled metallic nanorods at the far field by means of the scattering matrix (Jones matrix) of the system. 
In particular, we study the case of two coupled orthogonal metallic nanorods and derive a  formula for the far field 
circular dichroic response (CD) of the system that depends on two different phase factors  besides the electromagnetic coupling coefficient. We observe that the phase due to the chiral geometry is doubled and we show that the CD response of the system can be much greater for weakly coupled particles. 


\section{The scattering matrix }

Nanorods are the basic elements of a class of more complex systems. Their optical response can be modeled as oriented dipoles with polarization characteristics similar to that of linear polarizers in a certain interval of photon energy. We assume that the polarizability of each rod is fully anisotropic, i.e., it can polarize only along a particular direction.
Hence, scattering properties of a nanorod can be represented by a linear polarizer Jones matrix:

\begin{equation}
\mathbf{J}= \alpha  \left(
\begin{array}{cc}
\cos^2\theta & \cos\theta\sin\theta \\
\cos\theta\sin\theta & \sin^2\theta \\
\end{array}
\right)
\end{equation}
where $\alpha$
is the Lorentzian polarizability associated with the particle and $\theta$ is the orientation angle in the $x$-$y$ plane. 

\begin{figure}[!h]
\centering
\includegraphics[width=300pt,height=250pt]{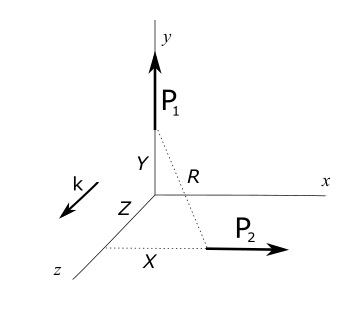}
\caption{Coupled dipoles (nanorods). CD response is maximum for weak coupling. }
\label{bem}
\end{figure}

We study a chiral configuration of coupled oriented dipoles (nanorods) depicted in Fig.\ref{bem}. Particles are  excited by a plane wave that propagates along the $z$-axis.  We calculate the components of the far field electric vector by taking into account the mutual interactions as described in the Appendix. 

\begin{equation}\label{P2}
E_{x}=\frac{F(e_2\varepsilon\alpha E_x+e_1\varepsilon\alpha^2\delta E_y)}{1-e_1^2\alpha^2\delta^2}  \end{equation}

\begin{equation}\label{P1} E_{y}=\frac{e_2F(e_1e_2\varepsilon\alpha^2\delta E_x+\varepsilon\alpha E_y)}{1-e_1^2\alpha^2\delta^2}
\end{equation}
where $\varepsilon$ is the permittivity of the medium, $E_{0x}$ and $E_{0y}$ are electric field components at  $z=0$, $F$ is the far field factor,
$e_1= e^{2\pi iR/\lambda}$, $e_2=e^{2\pi iZ/\lambda}$, $\delta$ is the coupling coefficient defined as $\delta=-XYk^2B/R^2$ ($B$ is given in the Appendix). There is an extra  $e_2$ in $E_y$  because the vertical rod is behind the horizontal rod at a distance $Z$.

From the far field components we extract the Jones matrix of the system:
\begin{equation}\label{J1}
\mathbf{J}=\frac{\varepsilon\alpha F}{1-e_1^2\alpha^2\delta^2}\begin{pmatrix}e_2&e_1\alpha\delta\\e_1e_2^2\alpha\delta&e_2\end{pmatrix}
\end{equation}
This is an asymmetric Jones matrix with a non-zero $\gamma$ parameter which is associated with the circular anisotropy of the system \cite{VMS}:
\begin{equation}
\gamma=\frac{i(J_{12}-J_{21})}{2}
\end{equation}
where $J_{ij}$ are the elements of the Jones matrix. According to Eq.\eqref{J1}:
\begin{equation}
\gamma=\frac{\varepsilon\alpha F (1-e_2^2)e_1\alpha\delta}{2(1-e_1^2\alpha^2\delta^2)}
\end{equation}
We also calculate the circular dichroic (CD) response of the system directly from Eq.\eqref{J1}. \begin{equation}\label{delta2}
\Delta I(R, Z, \lambda)=2i(-J_{11}J^*_{12}+J_{12}J_{11}^*-J_{21}J^*_{22}+J_{22}J^*_{21})
\end{equation}
where  $I_{RCP}$, $I_{LCP}$ are the scattering intensities corresponding to right- and left-handed circular polarization, and $\Delta I(R, Z, \lambda)= I_{RCP}-I_{LCP}$  \: is the differential scattering intensity which quantifies the circular dichroic response.
By using the property $J_{11}=J_{22}$, Eq.\eqref{delta2} simplifies to
\begin{equation}
\Delta I(R, Z, \lambda)=-4\mathbf{Im}(J_{11}^*(J_{12}-J_{21}))
\end{equation}
In terms of the elements of the Jones matrix given in Eq.\eqref{J1}

\begin{equation}\label{faz2}
\Delta  I(R,Z,\lambda)=8gg^*\sin{\bigg(\frac{2\pi Z}{\lambda}\bigg)}\mathbf{Re}\big(\alpha\delta e^{i2\pi R/\lambda}\big)
\end{equation}

\noindent where $g=\varepsilon\alpha F/(1-e_1^2\alpha^2\delta^2)$.


\section{Weak coupling}

In plasmonic models optical activity emerges due to the coupling. As shown in the previous section, the optical activity is directly proportional to the coupling coefficient. 
When the nanorods are well separated from each other $\gamma$ and $\Delta I$ vanishes, and one may expect that chiroptical effects would increase with decreasing distance.  However, according to Eq.\eqref{faz2}, for $Z=0$, the Jones matrix of the system reduces to a symmetric matrix, hence, there is no chiroptical effect for scattering in the $z$-direction \footnote{For $Z=0$, $\Delta I\neq 0$ in other scattering directions \cite{Pakizeh}}. In short,  when $Z=0$; $\gamma=0$,\: $\Delta I=0$ and, in general, $\Delta I$ may not be maximum for too closely packed nanorods. 
It is worth noting that, there are two phase factors in Eq.\eqref{faz2}. $e_2=e^{i2\pi Z/\lambda}$ is the phase due to the 3D chiral geometry and $e_1=e^{i2\pi R/\lambda}$ is the phase that involved in the dipole-dipole interaction. Only the imaginary part of $e_2$ appears in the equation. 
Maximum of the CD response is determined by the trade off between these two phase factors.

\begin{figure}[!h]
\centering
\includegraphics[width=300pt,height=250pt]{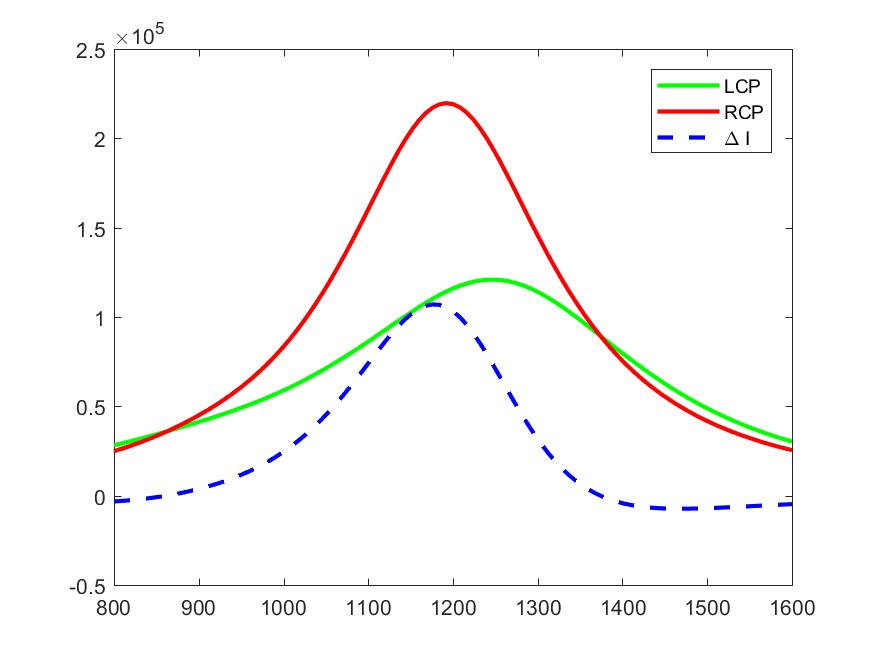}
\caption{Scattering intensities $I_{RCP}$, $I_{LCP}$ and $\Delta I$ for nanorods of length 400 nm and radius 50 nm with  $X=Y=430$ nm, $Z=240$ nm.}
\label{deltaI}
\end{figure}

We continue our work with the simulations of the  far field scattering intensities  $I_{RCP}$ and $I_{LCP}$ corresponding to $RCP$ and $LCP$ excitation polarization states. 
We observe that it is possible to maximize $\Delta I$ by playing with the spatial parameters associated with $e_1$ and $e_2$. Especially, for nanorods with length $L>$200 nm, $R$ can be made very large compared to the size of the rods (weak coupling). As an example, for nanorods with length 400 nm and radius 50 nm, by setting $X=Y=430$ nm and $Z=240$ nm ($R=$ 654 nm) $\Delta I$ is 1/3 of the total intensity at $\lambda_{PR}$ (plasmon resonance wavelength), i.e., $(I_{RCP}-I_{LCP})/(I_{RCP}+I_{LCP})\approx1/3$. BEM simulations for this configuration is given in Fig.\ref{deltaI}. Dashed line is for $\Delta I$. In the MATLAB implementation of the BEM method \cite{Hoh} we use optical constants of Au by Johnson and Christy \cite{Johns} (Supplemental Material).

\section{Conclusion}

In the case of coupled nanoparticles in three-dimensional space,
it is often necessary to study different excitation polarizations in different scattering directions and
matrix methods using all possible modes of excitation become important. In this work, we examine
the optical response of coupled metallic nanorods at the far field by means of the Jones matrix of the system. We study the case of two coupled orthogonal metallic nanorods in detail and derive a  formula for the far field 
circular dichroic response of the system that depends on two different phase factors  besides the electromagnetic coupling coefficient. We observe that the phase due to the chiral geometry is doubled and the CD response of the system can be much greater for weakly coupled particles.


\section{Appendix}

When a nanorod is excited by a plane wave the induced electric dipole moment vector, $\mathbf{P}$, is proportional to the incident electric field, $\mathbf{E}_0(\mathbf{r})$:
\begin{equation}
\mathbf{P}=\varepsilon\mathbf{J}\mathbf{E}_0(\mathbf{r}),
\end{equation}
where $\varepsilon$ is  the permittivity of the medium at the dipole (nanorod) position and $\mathbf{J}$ is the $2\times 2$ Jones matrix of the nanorod. 

When we put two nanorods close to each other we have to consider mutual interaction contributions. Each one of the dipoles experiences the field of the
other dipole which should be taken into account to find the actual dipole fields \cite{Albella}: 
\begin{subequations}
\begin{equation}\label{dipole1}
\mathbf{P}_{1}= \mathbf{J}_{1}[\varepsilon 
   \mathbf{E_o}(\mathbf{r}_{1}) + k^2 \mathbf{\bar{\bar{G}}}(\mathbf{r}_{1}-\mathbf{r}_{2})\cdot \mathbf{P}_{2}],
\end{equation}
\begin{equation}\label{dipole2}
\mathbf{P}_{2}= \mathbf{J}_{2}[\varepsilon 
   \mathbf{E_o}(\mathbf{r}_{2}) + k^2 \mathbf{\bar{\bar{G}}}(\mathbf{r}_{2}-\mathbf{r}_{1})\cdot \mathbf{P}_{1}],
\end{equation}
\end{subequations}
where $k$ is the wavenumber, $\mathbf{J}_{1}$, $\mathbf{J}_{2}$ are the Jones matrices of individual nanorods and $\mathbf{\bar{\bar{G}}}$ is the free-space electric dyadic Green's function with the following effect on a dipole vector: 

\begin{equation}
\mathbf{\bar{\bar{G}}}\cdot \mathbf{P} =\frac{1}{4\pi R}\bigg[\left(1 + \frac{i}{kR} - \frac{1}{k^2R^2}\right) \mathbf{P}  \\ + \left(-1 - \frac{3i}{kR} + \frac{3}{k^2R^2}\right)(\mathbf{\hat{u}}\cdot\mathbf{P})\mathbf{\hat{u}}\bigg],
\end{equation}
where $R$ is the distance and $\mathbf{\hat{u}}$ is the unit vector between the center of masses of particles. The notation can be simplified if we let,
\begin{subequations}
\begin{equation}
A=\frac{1}{4\pi R}\left(1 + \frac{i}{kR} - \frac{1}{k^2R^2}\right),
\end{equation}
\begin{equation}
B= \frac{1}{4\pi R}\left(-1 - \frac{3i}{kR} + \frac{3}{k^2R^2}\right),
\end{equation}
\end{subequations}
thus,
\begin{equation}
\mathbf{\bar{\bar{G}}}\cdot \mathbf{P} = A \mathbf{P} + B(\mathbf{\hat{u}}\cdot\mathbf{P})\mathbf{\hat{u}}.
\end{equation}

\begin{figure}
\centering
\includegraphics[width=300pt,height=250pt]{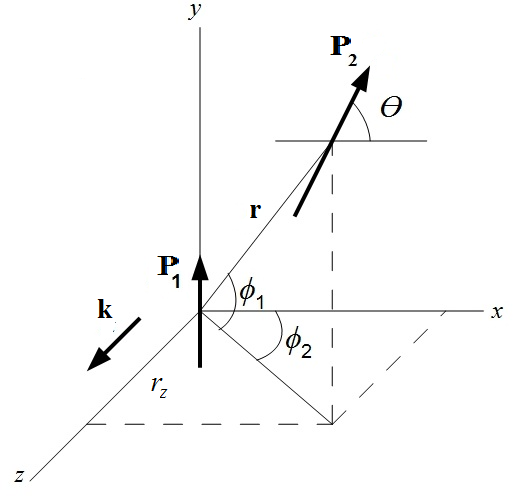}
\caption{Two oriented dipoles (nanorods)}
\label{geometry}
\end{figure}

We study the circular polarization effects for the geometry given in Fig.\ref{geometry} where  $\mathbf{r}$ ($|\mathbf{r}|=R$)  is the relative position vector between the dipoles. The plane wave excites $\mathbf{P_1}$ first and excites $\mathbf{P_2}$ after a delay.
According to Fig.\ref{geometry} $e_1 = e^{i2\pi R/\lambda}$ is the phase difference between the dipoles
along the distance $R$ and $e_2 = e^{i2\pi Z/\lambda}$ is the phase difference in the $z$-direction.
Jones matrix of the first dipole is fixed along the $y$ axis: 

\begin{equation}
\mathbf{J}_1= \alpha_{1}  \left(
\begin{array}{cc}
0 & 0 \\
0 & 1 \\
\end{array}
\right)
\end{equation}
The second dipole is tilted at an angle of $\theta$:

\begin{equation}
\mathbf{J}_2= \alpha_{2}  \left(
\begin{array}{cc}
a & b \\
b & c \\
\end{array}
\right)
\end{equation}
where $\alpha_1$ and $\alpha_2$ are the Lorentzian polarizabilities of the dipoles and
$a=\cos^2\theta$,
$b=\cos\theta\sin\theta$,
$c=\sin^2\theta$.
Let $C_1=\cos\phi_1, S_1=\sin\phi_1, C_2=\cos\phi_2, S_2=\sin\phi_2$ then 
the unit vector along $\mathbf{r}$ can be written as
\begin{equation}
\mathbf{\hat{u}}(\mathbf{r_2}-\mathbf{r_1})=(C_1C_2, S_1, C_1S_2)
\end{equation}

We calculate $\mathbf{P_1}$ and $\mathbf{P_2}$ with the Green function contributions: 
\begin{equation}\label{dipole1a}
\mathbf{P}_1=\varepsilon\mathbf{J}_1\begin{pmatrix}
E_{0x}\\E_{0y}
\end{pmatrix}
+k^2\mathbf{J}_1\begin{pmatrix}
e_1AP_{2x}+(C_1C_2P_{2x}+S_1P_{2y})C_1C_2e_1B\\e_1AP_{2y}+(C_1C_2P_{2x}+S_1P_{2y})S_1e_1B
\end{pmatrix}
\end{equation}

\begin{equation}\label{dipole2a}
\mathbf{P}_2=\varepsilon\mathbf{J}_2\begin{pmatrix}
e_2E_{0x}\\e_2E_{0y}
\end{pmatrix}
+k^2\mathbf{J}_2\begin{pmatrix}
C_1C_2S_1e_1BP_{1y}\\e_1AP_{1y}+S_1^2e_1BP_{1y}\end{pmatrix}
\end{equation}
$E_{0x}, E_{0y}$ are the components of the planewave excitation at $z=0$. 
We solve the components of the dipoles at the far field for scattering in the $z$-direction and we find the scattering matrix (Jones matrix) of the interacting system:

\begin{equation}\label{general}\begin{split}
\mathbf{J}=\frac{\varepsilon F}{N}\left[ e_2\alpha_1\begin{pmatrix}
0&0\\0&1\end{pmatrix}+ e_2\alpha_2\begin{pmatrix}
a&b\\b&c
\end{pmatrix}+e_1\alpha_1\alpha_2\begin{pmatrix}
0&\Delta_1 \\e_2^2\Delta_1&(1+e_2^2)\Delta_2\end{pmatrix} \right]
\end{split}
\end{equation}
where 
$N = 1- e_1^2
\alpha_1\alpha_2(2b\delta_1\delta_2 + c\delta_1^2 + a\delta_2^2)$, $F$ is the far field factor, $\delta_1=k^2(A+S_1^2B)$, $\delta_2=k^2(C_1C_2S_1B)$, $\Delta_1=b\delta_1+a\delta_2$ and $\Delta_2=c\delta_1+b\delta_2$ are the coupling coefficients that result from the dipole-dipole interaction. Here we write the Jones matrix of the system as a linear combination of three Jones matrices, first two of them corresponding to symmetric linear polarizer Jones matrices of individual (non-interacting) dipoles and the third one is an asymmetric Jones matrix due to the coupling coefficient and phase ($e_2$). All elements of the interaction Jones matrix are scaled by coupling coefficients which are functions of the distance between the dipoles so that for distant particles this coupling term consistently vanishes.

As a special case we study a simpler geometry given in Fig.\ref{bem} where we let $a=1, b=0, c=0$, $\alpha_1=\alpha_2=\alpha$, with $\Delta_1=\delta_2=\delta$ and $\Delta_2=0$.
Eq. \eqref{general} reduces to the following Jones matrix:
\begin{equation}\label{ozel}
\mathbf{J}=g\bigg[e_2\begin{pmatrix}1&0\\0&0\end{pmatrix}+e_2\begin{pmatrix}0&0\\0&1\end{pmatrix}+e_1\begin{pmatrix}0&\alpha\delta\\e_2^2\alpha\delta&0\end{pmatrix}\bigg]=g\begin{pmatrix}e_2&e_1 \alpha\delta\\e_1e_2^2\alpha\delta&e_2\end{pmatrix}
\end{equation}
where
\begin{equation}
g=\frac{\varepsilon\alpha F}{1-e_1^2\alpha^2\delta^2}
\end{equation}
Extremum points of  the denominator of the overall factor $g$ determines the intensity peaks corresponding to the hybridized modes which occur at the energies  that make $\mathbf{Re}(e_1\alpha\delta)=\pm 1$ \cite{PRB}. Separation between the higher and lower energy modes decreases and eventually they overlap for large $R$, but two modes can still be monitored by means of the parameter $e_1\alpha\delta$ which can be found from the Jones matrix.


\end{document}